\documentclass[a4paper]{article}

\usepackage{amsmath,amsfonts,amssymb,amsthm}

\newtheorem{defi}{Definition}
\newtheorem{teo}{Theorem}

\newtheorem{prop}[teo]{Proposition}

\newtheorem{rmk}{Remark }

\renewcommand{\aa}{a_{0}}

\newcommand{\dok}{c_N}

\begin{document}
	\title{Superintegrability and geometry: a review of the extended Hamiltonian approach}
	\author{Claudia Maria Chanu, Giovanni Rastelli\\ \\ Dipartimento di Scienze Umane e Sociali, \\ Universit\`{a} della Valle d'Aosta, Universit\'e de la Valle\'e d'Aoste\\ \\ Dipartimento di Matematica, Universit\`{a} di Torino,\\
		email: c.chanu@univda.it; giovanni.rastelli@unito.it}
	\maketitle

    \begin{abstract}
        We review the results of several of our papers about the procedure of {\it extension of Hamiltonians}, allowing the construction of families of superintegrable systems with 
        non-trivial polynomial first integrals (or symmetry operators) of arbitrarily high degree. In particular, we focus on the geometric structures involved by the procedure: warped manifolds and Riemannian coverings. Examples of superintegrable systems, classical and quantum,  with the structure of extended Hamiltonians are anisotropic harmonic oscillators, Kepler-Coulomb, Tremblay-Turbiner-Winternitz and Post-Winternitz systems.
    \end{abstract}

\maketitle

\section{Introduction}

Hamiltonians of finite dimension $n$ are superintegrable when they admit more than $n$ functionally independent first integrals. The maximum number of functionally independent first integrals is $2n-1$. For finite dimensional quantum systems, algebraic  independence takes the place of functional independence: $m$ symmetry operators are algebraically independent when none of them can be written as a polynomial, with complex coefficients,  of the other $m-1$ (see for example \cite{MPW} and references therein).

When the symmetry operators are obtained as Laplace-Beltrami quantization of polynomial in the momenta constants of motion, functional independence of the last ones implies algebraic independence of the first ones \cite{MPW}.

Consequences of maximal superintegrability ($2n-1$ independent constants of motion) are, for the classical systems, closure of bounded orbits and, for the quantum ones, complete degeneration of the energy levels.

Quadratical superintegrability (when all the independent first integrals are quadratic in the momenta, or the symmetry operators are all of second order) is in many cases   a consequence of multiseparability: the Hamilton-Jacobi  (Schr\"odinger) equation admits solutions additively (multiplicatively) separated in several different coordinate systems. For example, the Kepler-Coulomb system. In certain cases, however,  multiseparability is not enough to grant maximal superintegrability, for example the Calogero or Calogero-Moser system \cite{BCRCal}.

In this case, as in many other, polynomial first integrals of degree greater than two are necessary for maximal superintegrability.

A modern systematic  research about superintegrability involving first integrals, or symmetry operators, of high order was undertaken in 1960' by Smorodinsky, Winternitz and collaborators. 

Several papers have been published about superintegrability in the last two decades, a short  historical review of the modern approach to superintegrable systems, up to 2013,  can be found in \cite{MPW}. Since then, many articles appeared, devoted to superintegrable Hamiltonian systems. For example, focusing on the algebraic strucure \cite{s1,s2,s3}, on superintegrability of St\"ackel separable systems \cite{s4,s5,s6,s7}, on the geometry of the configuration manifold \cite{s8,s9,s10,s11,s12}, higher order superintegrability \cite{s13,s14,s15,Bo}, superintegrability of systems with electro-magnetic fields \cite{s16,s17}.


In this review article, we summarize, putting them into a unified perspective, several articles we published, many of them with L. Degiovanni, about superintegrability in Hamiltonian systems, where the existence and construction  of non trivial first integrals of high degree is the main topic.

It is well known that the Jacobi-Calogero system of three points $(x_i)$ on a line, with potential
\begin{equation}
V=(x_1-x_2)^{-2}+(x_2-x_3)^{-2}+(x_3-x_1)^{-2},
\end{equation}
is maximally superintegrable, with four integrals quadratic in the momenta and one cubic. All the quadratic ones are associated with the multiseparability of the system, in five different coordinate systems of $\mathbb E^3$ \cite{BCRCal}. These are five linearly independent quadratic first integrals, but only four of them are functionally independent. Written in any one of those coordinates, the Jacobi-Calogero potential becomes
\begin{equation}
V=\frac l{r^2\sin ^2 3\phi}, \quad l\in \mathbb R,
\end{equation}
where $(r,\phi)$ are polar coordinates in the plane.
In \cite{CDR0} it is shown as the cubic first integral of the system can be obtained by analyzing the system in these coordinates. It emerged that the existence of the cubic integral is strictly related to the number 3 appearing in $V$ as a factor of the ignorable coordinate $\phi$. After changing 3 into other integers, or their reciprocals, new polynomial first integrals were obtained. Naturally, the corresponding Hamiltonians are no longer simply related with systems of points on a line, or other similar physical configurations. 

We will see later on  an interpretation of these systems in terms of Hamiltonian systems on Riemannian coverings.

These modified Jacobi-Calogero systems were generalized in \cite{TTW} as a modification of the Smorodinsky-Winternitz system. 
The Tremblay-Turbiner-Winternitz (TTW) Hamiltonian can be written as \cite{TTW,TTWcdr} 
\begin{equation}\label{TTW1}
H=\frac 12 p_r^2+\frac{1}{r^2}\left(\frac 12p_\Phi^2+\frac {\alpha_1}{\cos^2h\Phi}+\frac {\alpha_2}{\sin^2h\Phi}\right)+\omega r^2, \quad r>0, \quad 0\leq \Phi <2\pi,
\end{equation}
where $\alpha_i$ and $\omega$ are real constants, $h$ is a positive rational number and $H$ is    locally defined on the cotangent bundle of the Euclidean  plane. Its maximal superintegrability was conjectured in \cite{TTW}, and eventually proved in \cite{KM1}. In the TTW system, a generic rational parameter $h$ appears as factor of the ignorable coordinate $\phi$ as the number $3$ in the Jacobi-Calogero potential of above. The maximal superintegrability of the system holds for any $h$, and  a polynomial first-integral, or symmetry operator for the quantum counterpart, of order depending on $h$ is provided.

From the TTW system, via coupling constant metamorphosis, it can be obtained another superintegrable systems, known as Post-Winternitz (PW) \cite{PWccm, KMccm}, whose difference from the TTW system is essentially the presence of a Kepler-Coulomb scalar term instead of the harmonic oscillator one in the TTW. 

 In \cite{CDRPol} we developed, with L. Degiovanni, a general theory for classical superintegrable systems of TTW type, extended later to include coupling constant metamorphosis of TTW, the PW systems, and later a quantization scheme for all these systems \cite{CDRraz, TTWcdr,CDRpw}.

The main result of this research is the determination of a wide family of Hamiltonian systems admitting non-trivial first integrals, or symmetry operators, of arbitrarily high order in a variety of Riemannian manifolds, with different curvatures and  signatures. Such a family of systems, that we called {\it extension of Hamiltonian systems} or {\it extended Hamiltonians} can be employed for the analysis of integrability and superintegrability on curved manifolds and of different procedures of quantization \cite{Rbj, CDRmlbq, CRsl}.

Written in polar coordinates, the TTW system  shows its structure of warped Hamiltonian, based on a warped Riemannian manifold. Such a structure is common with all the systems obtained by us as extended Hamiltonians \cite{CDRfi,CDRgen,CDRraz}, and allows the explicit determination of additional polynomial first integrals of high degree. In \cite{coverings} it is shown how the first integrals obtained by this way are actually always globally defined, differently from those obtained by other approaches. Indeed, all the systems of the TTW family, as well as any extended Hamiltonian system, can be considered as induced on  Riemannian coverings of the Riemannian manifold corresponding to the value $k=1$ of the parameter, where, with respect to (\ref{TTW1}), $k=1/h$. These coverings, when $k^2<1$,  are known as conic manifolds, or footballs, when based on Euclidean spaces or spheres respectively. With the help of  this interpretation, issues of global definition for the first integrals, or for the symmetry operators, for $k$ not integer become evident. In \cite{Prague} it is shown how the degeneration of the energy levels for the Kepler-Coulomb and for the Klein-Gordon equation in Schwarzschild background is affected by interpreting them on Riemannian coverings. We recall that the Kepler-Coulomb system  results as a particular case  of the PW system.

\section{Definitions and relevant properties}\label{s2}
Let $L(q^i,p_i)$ be a Hamiltonian with $N$ degrees of freedom, that is defined on the cotangent bundle of an $N$-dimensional manifold.
We say that $L$ {\em admits extensions}, if there exists $(c,c_0)\in \mathbb R^2
- \{(0,0)\}$ such that there exists a non-null solution $G(q^i,p_i)$ of
\begin{equation}\label{e1}
X_L^2(G)=-2(cL+c_0)G,
\end{equation}
where $X_L$ is the Hamiltonian vector field of $L$.
If $L$ admits extensions, then, for any $\gamma(u)$ solution of the ODE
\begin{equation}\label{eqgam}
\gamma'+c\gamma^2+C=0,
\end{equation}
depending on the arbitrary constant parameter 
$C$,	
we say that any Hamiltonian 
$H(u,q^i,p_u,p_i)$ with $N+1$ degree of
freedom of the form
\begin{equation}\label{Hest}
H=\frac{1}{2} p_u^2-k^2\gamma'L+ k^2c_0\gamma^2
+\frac{\Omega}{\gamma^2},
\qquad k\in \mathbb{Q}-\{0\}, \   \Omega\in\mathbb{R}
\end{equation}
is an {\em  extension of $L$}.

Extensions of Hamiltonians where introduced
in \cite{CDRfi} and studied because they admit polynomial in the momenta first integrals generated via a recursive algorithm. Moreover, the degree of the first integrals is related with the (choice of) numerator and denominator of the rational parameter $k=m/n$, $m,n\in \mathbb{N}^*$.
Indeed, for any $m,n\in \mathbb N-\{0\}$, 
let us consider the operator
\begin{equation}\label{Umn}
U_{m,n}=p_u+\frac m{n^2}\gamma X_L
\end{equation}

\begin{prop}\cite{CDRraz}
For $\Omega=0$,
the Hamiltonian (\ref{Hest}) 
is in involution with the function
	\begin{equation}\label{mn_int}
	K_{m,n}=U_{m,n}^m(G_n)=\left(p_u+\frac{m}{n^2} \gamma(u)   X_L\right)^m(G_n)
	\end{equation}
	where $m/n=k$ and $G_n$ is the $n$-th term of the recursion
	\begin{equation}\label{rec}
	G_1=G, \qquad G_{n+1}=X_L(G)\,G_n+\frac{1}{n}G\,X_L(G_n), 
	\end{equation}
starting from any solution $G$ of (\ref{e1}).
\end{prop}

\begin{rmk}  The  functions $G_n$ satisfy
	$$
	X_L^2G_n=-2n^2 (cL+c_0)G_n.
	$$
\end{rmk}

For $\Omega\neq 0$, the recursive construction of a first integral is more complicated:  we consider the following function, depending on two non-zero integers
$2s,r$ (the first one has to be even)
	\begin{equation}
	\bar K_{2s,r}=\left(U_{2s,r}^2+2\Omega \gamma^{-2}\right)^s(G_r).\label{ee2}
	\end{equation}
where 
the operator $U^2_{2s,r}$ is defined {according to
	(\ref{Umn})} as
$$
U^2_{2s,r}=\left( p_u+\frac{2s}{r^2} \gamma(u)   X_L
\right)^2,
$$
and 
$G_r$ is, as in (\ref{mn_int}),  the $r$-th term of the recursion (\ref{rec}).  
with $G_1=G$ solution of (\ref{e1}).
For $\Omega=0$ the functions (\ref{ee2})
reduces to (\ref{mn_int}) and thus can be computed 
also when the first of the indices is odd. 

We have 

\begin{teo}\cite{TTWcdr}
For any $\Omega\in \mathbb{R}$,
the Hamiltonian (\ref{Hest}) with $k=m/n$ 
satisfies
	for  $m=2s$, 
		\begin{equation}\label{cp}
		\{H,\bar K_{m,n}\}=0,
		\end{equation}
for $m=2s+1$,
		\begin{equation}
		\label{cd}
		\{H ,\bar K_{2m,2n}\}=0.
		\end{equation}
\end{teo}

We call $K$ and $\bar{K}$, of the form (\ref{mn_int}) and
(\ref{ee2}) respectively, \emph{ characteristic first integrals}  of the corresponding extensions

\begin{rmk}

In \cite{CDRgen} it is proven that the ODE (\ref{eqgam}) defining $\gamma$ is a necessary condition in order to get a characteristic first integral of the form (\ref{mn_int}) or
(\ref{ee2}). 
According to the value of $c$, the explicit form of $\gamma(u)$ is given (up to constant translations of $u$) by
\begin{equation}\label{fgam}
\gamma= \left\{ 
\begin{array}{lc}
-C u & c=0 \\
\frac{1}{T_\kappa (cu)}=\frac{C_\kappa (cu) }{S_\kappa (cu)}
& c\neq 0
\end{array}
\right.
\end{equation}
where $\kappa=C/c$ is the ratio of the constant parameters appearing in (\ref{eqgam}) and $T_\kappa$, $S_\kappa$ and $C_\kappa$  are the trigonometric tagged functions
 \cite{7ter}
		$$
		S_\kappa(x)=\left\{\begin{array}{ll}
		\frac{\sin\sqrt{\kappa}x}{\sqrt{\kappa}} & \kappa>0 \\
		x & \kappa=0 \\
		\frac{\sinh\sqrt{|\kappa|}x}{\sqrt{|\kappa|}} & \kappa<0
		\end{array}\right.
		\qquad
		C_\kappa(x)=\left\{\begin{array}{ll}
		\cos\sqrt{\kappa}x & \kappa>0 \\
		1 & \kappa=0 \\
		\cosh\sqrt{|\kappa|}x & \kappa<0
		\end{array}\right.,
		$$
		$$
		T_\kappa(x)=\frac {S_\kappa(x)}{C_\kappa(x)},
		$$
(see also \cite{CDRraz} for a summary of their properties).
\end{rmk}

\begin{defi}
Given $n$ Riemannian manifolds $(Q_k, ds_k^2)$,  $(k=1,\ldots,n)$,  their warped product is the Riemannian manifold $Q=\times _{k=1}^nQ_k,$ with metric $ds^2= \alpha^k ds_k^2$, where  $\alpha^k:Q_1
\rightarrow \mathbb R$ or $\mathbb C$ for \(k=2,\ldots,n\) and \(\alpha^1 = 1\). 
\end{defi}

A simple example of non-trivial warped product of Riemannian manifolds is  the  Euclidean plane  with metric $ds^2=dr^2+r^2d\theta^2$ in polar coordinates $(r,\theta)$ which can be interpreted as warped product of  $(\mathbb R^1)^+\times \mathbb S^1$ with $\alpha^1=1$, $\alpha^2=r^2$. 

Given  a Riemannian manifold, the possibility of writing it as warped product of lower dimensional manifolds is characterized by several theorems involving its geodesic completeness, curvature, etc.\ (see for example \cite{Ta}). 

The concept of warped product can be straightforwardly generalized to twisted product of Hamiltonian systems \cite{CDRGhent,CRbs}:
\begin{defi}
Let $H_k$ 
be $n$ non-degenerate Hamiltonians on $T^*Q_k$ with canonical coordinates $(q^{i_k},p_{j_k})$. Let $\alpha^k$ be $n$ functions defined on $Q=\times_{k=1}^n Q_k\rightarrow \mathbb R$ (we do not consider here the more general case of $\alpha_k$ defined on the whole $T^*Q$). Let 
\begin{equation}
H=\alpha^kH_k
\end{equation}
be a Hamiltonian in $\times_{k=1}^n T^*Q_k=T^* Q$  with the obvious canonical coordinates.

Then, $H$ is the twisted product of the Hamiltonians $H_k$. 
\end{defi}

The concept of twisted product of Hamiltonians is important to understand block-separation of variables, a weaker form of St\"ackel separation analyzed in \cite{CRbs}.

\begin{rmk}
    The metric of extended Hamiltonians is the metric of a warped manifold and the extended Hamiltonian is the warped product of  Hamiltonians

    \begin{eqnarray}\label{2.12}
        H=\alpha^1\left(\frac{1}{2} p_u^2+ k^2c_0\gamma^2+
\frac{\Omega}{\gamma^2}\right)+\alpha^2 L,
\quad \alpha_1=1, \alpha_2=-k^2\gamma'.
\end{eqnarray}
\end{rmk}

\section{Superintegrability}

It is proved in \cite{CDRfi,TTWcdr} that the  characteristic first integrals $K$ or $\bar K$  are functionally independent from $H$, $L$, and from any 
first integral $I(p_i,q^i)$ of $L$.
This means that the extensions of (maximally) superintegrable Hamiltonians	are 
(maximally) superintegrable Hamiltonians with one additional degree of freedom. The possibility of obtaining in this way new superintegrable systems from old ones has been investigated in \cite{CDRsuext}, where the superintegrable extensions of several quadratically superintegrable Hamiltonians on constant curvature manifolds are obtained. Another application of this property of extended Hamiltonians is made in \cite{CDRmlbq}, where a recursive family of superintegrable extended Hamiltonians is analyzed, together with its modified Laplace-Beltrami quantization.


The explicit expression of the characteristic first integrals is given as follows \cite{TTWcdr}.
For $r\leq m$, 
we have
\begin{equation}\label{EEcal}
U_{m,n}^r(G_n)= P_{m,n,r}G_n+D_{m,n,r}X_{L}(G_n),
\end{equation}
with
 \begin{equation*}
G_n=\sum_{ j=0}^{[(n-1)/2]}\binom{n}{2 j+1}\, G^{2 j+1}(X_L G)^{n-2 j-1}(-2)^j(cL+c_0)^ j,
 \end{equation*}
 \begin{equation*}
P_{m,n,r}=\sum_{ j=0}^{[r/2]}\binom{r}{2 j}\, \left(\frac mn  \gamma \right)^{2 j}p_u^{r-2 j}(-2)^j(cL+c_0)^ j,
 \end{equation*}
 \begin{equation*}
D_{m,n,r}=\frac 1{n}\sum_{ j=0}^{[(r-1)/2]}\binom{r}{2 j+1}\, \left(\frac mn  \gamma \right)^{2 j+1}p_u^{r-2 j-1}(-2)^j(cL+c_0)^ j, \quad m>1,
 \end{equation*}
where $[\cdot]$ denotes the integer part and $D_{1,n,1}=\frac 1{n^2} \gamma$.

The expansion of the first integral (\ref{ee2}) is
\begin{equation*}
\bar K_{2m,n}=\sum_{j=0} ^{m}\binom{m}{j}\left(\frac {2\Omega}{\gamma^2}\right)^jU_{2m,n}^{2(m-j)}(G_n),
\end{equation*}
with $U^0_{2m,n}(G_n)=G_n$.
 
\begin{rmk} 
If $L$ is a natural Hamiltonian with a metric and a potential which are algebraic (resp. rational) functions of the $(q^i)$ and if $L$ admits and extension associated with a function $G(q^i,p_i)$ which is polynomial in the momenta and algebraic (resp. rational) in the ($q^i$), then the extension of $L$ with $\gamma=(cu)^{-1}$ has first integrals which are polynomial in the momenta and algebraic (resp. rational) in the $(q^i)$.
The first integrals $\bar K_{2m,n}$ are polynomial w.r.t. $G$, $L$, $X_L G$, $\gamma$ and $p_u$. Indeed, by (\ref{e1}), it is easy to check that  $X_L(G_n)$ is a polynomial in  $G$, $X_L G$, $L$, as well as $G_n$.
\end{rmk}

\begin{rmk}
    The procedure of extension is not restricted to natural Hamiltonians $L$ only. In \cite{NoNN} it is shown how to obtain extended Hamiltonians from an Hamiltonian $L$ of degree four in the momenta, and from the two-point-vortices Hamiltonian.
\end{rmk}

 \section{Geometry 1: warped manifolds}
	
The equation (\ref{e1}) is the fundamental condition for the  extension procedure. When $L$ is a $N$-dimensional natural Hamiltonian with metric tensor $\mathbf g=(g_{ij})$ and $G$ does not depend on the momenta, (\ref{e1}) splits into several equations, the coefficients of monomials in the momenta equated to zero, whose leading term is the Hessian equation
\begin{equation}\label{hesseq}
\nabla_i\nabla_j G+cGg_{ij}=0,\quad i,j = 1,\ldots ,N.
\end{equation}
This equation is fundamental in theory of warped manifolds, see for example \cite{Ta},
where it is 
proved, for manifolds of positive definite signature, the  connection between the constant $c$, the curvature of the metric $g$ (Theorem 2, where it is also proved that, if the metric is geodesically complete and $c\neq0$, then $g$ is of constant curvature $c$) and the warped structure of $g$ whenever the equation (\ref{hesseq}) admits solutions $G(q^k)$ (Lemma 1.2 of \cite{Ta}). In few words, metrics admitting non-trivial solutions of the Hessian equation have a warped structure.

 By relaxing the hypothesis of the
 these statements, for example allowing the non geodesic completeness of the manifold in the case of $c\neq 0$, solutions of (\ref{hesseq}) can be found also in non-constant curvature manifolds, \cite{CDRmlbq}.
Moreover, the structure itself of the extensions
$H$ of natural Hamiltonians shows that the extended metric is again a warped metric.

Since the metric of an extended Hamiltonian is warped by construction, and the metric of $L$ must be warped in reason of (\ref{hesseq}), it is possible a recursive construction of Hamiltonian systems, starting from one-dimensional ones, each one being the extension of a lower-dimensional extension. An example of such recursion is provided in \cite{CDRmlbq}, where the procedure is iterated up to five dimensions, obtaining a recursion of maximally superintegrable geodesic systems. The Laplace-Beltrami quantization of the classical systems is obtained by determining suitable quantum corrections. The warped product of metrics can be naturally extended to warped products of Hamiltonians \cite{CDRGhent}.

\section{Intrinsic characterization}

In order to show that a natural Hamiltonian is an extended Hamiltonian, we have first to check if  it can be written, by a point-transformation of coordinates, in the form of a warped Hamiltonian (\ref{2.12}). We will use the intrinsic characterization of such natural Hamiltonians making a simple generalization  of the result given in \cite{CDRgen} (Theorem 9). The characterization has been applied in \cite{CRcs} to show the structure of extended Hamiltonian of an algebraic superintegrable system \cite{CS}.

Let us consider a natural Hamiltonian 
\begin{equation}\label{Hgen}
H=\frac 12 \tilde g^{ab}p_ap_b+\tilde V
\end{equation}
on a $(n+1)$-dimensional  Riemannian manifold $(\tilde Q, \tilde{\mathbf g})$ and let $X$ be a conformal Killing vector of 
$\tilde{\mathbf g}$, that is a vector field satisfying 
 \begin{equation*}
[X,\tilde{\mathbf g}]=\mathcal L_X \tilde{\mathbf g}= \phi \tilde{\mathbf g},
 \end{equation*}
where $\phi$ is a function on $\tilde Q$, called {\em conformal factor}, $\mathcal L_X$ is the Lie derivative with respect to the vector field $X$ and $[\cdot,\cdot]$  denotes the Schouten-Nijenhuis bracket. 
We define $X^\flat$ as the  1-form obtained by lowering the indices  of the vector field $X$ with the metric tensor $ \tilde{\mathbf g}$.

\begin{teo}
If on $\tilde Q$ there exists a conformal Killing vector field $X$ with conformal factor $\phi$ such that
\begin{eqnarray}
dX^\flat \wedge X^\flat=0, \label{Xnorm}\\
d\phi\wedge X^\flat=0, \label{phi_u}\\
d\left\|X\right\| \wedge X^\flat=0, \label{Xconst}\\
d(X(\tilde V)+\phi\tilde V)\wedge X^\flat=0, \label{Vok}
\end{eqnarray}
then, on \(\tilde Q\) there exist  coordinates $(u,q^i)$ such that 
$\partial_u$ coincides up to a rescaling with $X$ and the natural Hamiltonian (\ref{Hgen}) has the form 
\begin{equation}\label{Hext}
	H=\frac 12 p_u^2+f(u)+\left(\frac {m}n\right)^2\alpha(u) L, \qquad m,n\in \mathbb{N}-\{0\}.
	\end{equation}
Morover, if 
\begin{equation}
\tilde R(X)=aX, \quad a\in \mathbb{R}, \label{Xeig}
\end{equation}
where $\tilde R$ is the Ricci tensor of the Riemannian manifold $(\tilde Q, \tilde{\mathbf g})$,
then the function $\alpha(u)$ in (\ref{Hext}) satisfies the condition $\alpha=-\gamma'$ with
$\gamma$ solution of (\ref{eqgam}).
\end{teo}

\begin{rmk} 
	Conditions (\ref{Xnorm}), (\ref{phi_u}), (\ref{Xconst}) are the already known  intrinsic characterization of a warped metric,
	while condition (\ref{Vok}) intrinsically assures  that the scalar part of the Hamiltonian is compatible with the warped structure. It is remarkable the fact that an intrinsic condition -- Eq. \ref{Xeig} -- characterizes also the subset of the warped metrics which are involved in extended Hamiltonians.  
	The extended potentials with $\Omega=0$ are intrinsically characterized by $X(\tilde{V})+\phi\tilde{V}=0$.  
\end{rmk}

The TTW system is obtained in \cite{TTWcdr}, where polar coordinates are denoted by $(u,q)$ instead of $(r,\phi)$, from the Hamiltonian
\begin{equation}\label{TTW2}
H=\frac 12p_r^2+\frac1{k^2r^2}\left(\frac 12p_\phi^2+\frac{c_1+c_2\cos \phi}{\sin^2\phi}\right)+\omega r^2,
\end{equation}
by putting $k=\frac 1{2h}$, $\Phi=k\phi$ and
 \begin{equation*}
c_1=\frac{\alpha_1+\alpha_2}{2h^2},\quad c_2=\frac{\alpha_2-\alpha_1}{2h^2}.
 \end{equation*}

We see now that the Hamiltonian (\ref{TTW2}) is in the form of an Extended Hamiltonian \cite{TTWcdr}, for which polynomial first integrals of degree depending on $k\in \mathbb Q^+$ are explicitly determined via an algebraic-differential operator acting on
 \begin{equation*}
\frac 12p_\phi^2+\frac{c_1+c_2\cos \phi}{\sin^2\phi}.
 \end{equation*}
Therefore, we can consider the TTW system (\ref{TTW1}), which is equivalent to (\ref{TTW2}), as also equivalent  to
\begin{equation}\label{TTW3}
H=\frac 12p_r^2+\frac1{r^2}\left(\frac 12p_\Phi^2+\frac{\tilde c_1+\tilde c_2\cos \Phi}{\sin^2\Phi}\right)+\omega r^2,\quad \Phi=k\phi,
\end{equation}
where $\tilde c_i=\frac {c_i}{k^2}$, 
and we may consider (\ref{TTW2}) and (\ref{TTW3}) as  defined on a covering  manifold of the Euclidean plane, where (\ref{TTW2})  becomes (\ref{TTW3})  via the transformation $\Phi=k\phi$, with $0\leq \phi< 2\pi$.

The PW system can be written as the CCM of the TTW system. In \cite{CDRpw} it is shown that the CCM of an extended Hamiltonian, while not an extended Hamiltonian itself, admits a polynomial first integral, the CCM of the characteristic first integral, which is again generated by the powers of an operator, as for all extended Hamiltonians.

Therefore, it appears that the theory of extended Hamiltonians could be generalized to include the CCM of extended Hamiltonians, and possibly more general systems. At the moment, no research has been done in this direction.

\section{Geometry 2: Riemannian coverings}
The parameter $k^2$ introduced in extended Hamiltonians needs an interpretation. If we consider the simple case when $L$ is one-dimensional and its configuration manifold is a circle and $-\gamma'=u^{-2}$, as is the case of the TTW system, then we may introduce polar coordinates adapted to the warped structure of the extended manifold, which, for $k=1$,  coincides with the Euclidean plane minus the origin. Then, as done in \cite{TTW, MPW}, we can reparametrize the angular coordinate with the new variable $\Phi=k\phi$. As a consequence, the metric coincides now with the Euclidean metric for any value of $k$, but any trigonometric function in $L$, depending originally on $\phi$, will depend now on $\Phi=k\phi$. We are now facing the problem that, for certain values of $k$, some of the trigonometric functions in $L$ are no longer smooth and periodic on all the manifold, for example when $k$ is not integer. this behavior can be understood when we see that in this case the metric of the extended manifold becomes the pull back of the metric of the manifold with $k=1$, the Euclidean plane in this case. In this way, the metric of the extended manifold become the metric of a Riemannian covering of the Euclidean plane, where $k$ is the order  of the covering, i.e. the number of times that it covers the Euclidean plane. It is evident now that any positive integer value of $k$ does not raises problems with trigonometric functions in $L$, while non-integer ones, in general, do.

 Indeed, for $k$ not integer the first integrals of the TTW system obtained in \cite{TTW, MPW} and references therein, are not always  globally defined, while those determined via the extension procedure are globally defined for any rational value of $k$, since in this case they are determined without the reparametrization $\Phi=k\phi$. The same considerations hold for the Post-Winternitz system obtained from the TTW via coupling constant metamorphosis (CCM) \cite{Hi,PWccm}, written as the CCM of (\ref{TTW2}) in\cite{CDRpw} as
\begin{equation}\label{PW1}
H=\frac 12 p_r^2+\frac {k^2}{4r^2}\left( \frac 12 p_\phi^2+\frac {c_1+c_2 \cos \phi}{\sin^2 \phi}\right)-\frac E{2r},
\end{equation}
and the problem become relevant when the two systems reduce to the harmonic oscillator and the Kepler system respectively, systems that, via the transformation $\Phi=k\phi$ are defined on covering manifolds of the Euclidean plane. 

It happens that the characteristic first integrals of extended Hamiltonians never depend on variables parametrized by the parameter $k$, differently from the high degree first integrals determined by different techniques, so that they are not affected by topological problems. As a matter of fact, however, the determination of other first integrals on the extended phase space seems nevertheless affected by topology, for example, the Laplace vector for the Kepler system appearing in the PW system (\ref{PW1}).
For a detailed treatment of the topic, see \cite{coverings}.

\section{Quantization}

As long as the first integrals are all quadratic in the momenta, the problems of quantization are limited to the search of suitable scalar quantum corrections, in general determined by the geometry of the configuration manifold, as described in \cite{CDRmlbq}. The determination of the quantization rule for first integrals of higher degree in the momenta is less straightforward. Several different rules of quantization can be applied in this case, and the procedure of extensions can provide examples of irreducible first integrals of unlimited degree where to test the quantization rules. See for example \cite{Rbj} where the extension of the anisotropic oscillator allows to detect the different behavior of Born-Jordan and Weyl quantization rules on the maximally superintegrable structure of the system. A general quantization procedure for extended Hamiltonians is still lacking. However, for the important case of non-zero curvature extended metrics, a quantization procedure based on factorization of the operators, similar to creation and annihilation operators in the case of the harmonic oscillator, has been developed in \cite{CRsl}. The procedure produces quantum superintegrable systems from extended Hamiltonians on any n-dimensional non-flat manifold, with the following important characteristics: they are not entirely in polynomial form, they are not commuting for any wavefunction, but only for the eigenfunctions of the operator $\hat L$, corresponding to the Hamiltonian $L$ involved in the extension procedure. On the other side, the same limitations apply to the corresponding quantization of the harmonic oscillator. The following quantization is described into details in \cite{CRsl} and it is applicable to extended systems of TTW type on non-flat manifolds.

Starting from the classical case, we can  write a factorized characteristic first integral of any $H$  which is an extension
(\ref{Hest}) with $c\neq 0$ of a Hamiltonian $L$ admitting
  functions $G^\pm$ (ladder functions) such that
 \begin{equation}\label{gpm}
 X_LG^\pm=\pm\sqrt{-2(cL+c_0)}G^\pm,
 \end{equation}
where $X_L$ is the Hamiltonian vector field of Hamiltonian $L$ with respect to the canonical symplectic structure.

\begin{defi}
A function $F(q^i,p_i)$ is a  ladder function of $L$ if there exist a function $f(L)$ such that
\begin{equation}
    X_L(F)=f(L)F.
\end{equation}
\end{defi}

 Let us assume that $L$ is a $N$-dimensional natural Hamiltonian 
 \begin{equation}\label{L}
 L=\frac 12 g^{ij}p_ip_j +V(q^i).
 \end{equation}

 \begin{prop}
A ladder function of $L$
is given by
\begin{equation} 	\label{18}
G^\pm=\pm(\nabla G)^ip_i+ G f+ \frac 1f c_1,
\qquad {\textstyle where}\quad f=\sqrt{-2(cL+c_0)},
\end{equation}
with the function $G(q^i)$ and the constant  $c_1$ satisfying
\begin{align}
\nabla \nabla G+ c {\mathbf g}G=0,\label{HE1}\\
\nabla V \cdot \nabla G - 2(cV+c_0)G + c_1=0,\label{VE1}
\end{align}
where $[\cdot,\cdot]$ are the Schouten brackets.
\end{prop}

Remark that (\ref{HE1}) and (\ref{VE1}) coincide with the conditions (\ref{e1}) for $L$ to admit an extended Hamiltonian when $c_1=0$ and $G$ is independent from momenta.

After (\ref{eqgam}), we can write any extension (\ref{Hest}) of $L$ when $c\neq 0$ as
\begin{equation}\label{H}
H =\frac 12 p_u^2+\frac 12 \left(\gamma^2+\frac{C}{c}\right)M^2+\frac{\omega^2}{2\gamma^2}-\frac{C}{c} k^2c_0,
\end{equation}
where we set $M=k\sqrt{2(cL+c_0)}$
and $\omega^2=2\Omega$.

Therefore, we have 

\begin{prop}\label{teo}
	Given a Hamiltonian $L$ such that equation (\ref{gpm}) admits a solution $G^\pm$ for some constants $c$ and $c_0$ not both zero, then, 	
	(i) $L$ admits extensions, 
	(ii) for any extension $H$ (\ref{H}) of 
	$L$,  the functions
	$$
	X^\pm=(G^\pm)^{2n}(A^\mp)^{2m}(D^\pm)^m,
	$$
	with 
	$$
	A^\pm=\mp \frac i{\sqrt{2}} p_u+\frac{\omega}{\sqrt{2} \gamma} -\frac M{\sqrt{2}}\gamma,
	$$
	$$
D^\pm=\mp \frac i{\gamma} p_u+\frac{\omega}{\gamma^2} -\frac{H}{\omega}-
\frac{C}{c\omega}\left(c_0k^2- \frac{M^2}{2} \right),
	$$
are factorized first integrals of $H$.
\end{prop}

Our aim is to give a general method for constructing a $(N+1)$ quantum Hamiltonian operator admitting a factorized symmetry, starting from 
a natural Hamiltonian
$L$ with $N$ degrees of freedom which admits a ladder function  such that there exists a non-trivial solution of  
(\ref{HE1}) and (\ref{VE1}).
As a first step, we look for shift and ladder operators similar to those given in \cite{CKNd}.
Finally, as in \cite{CKNd}, we use them in order to construct a symmetry operator for the extended quantum Hamiltonian.
The structure of this Hamiltonian operator
is strictly related with the classical extension  of $L$,
but does not exactly coincide with the Laplace-Beltrami quantization of an extension of
$L$ \cite{CDRmlbq}.

\begin{defi} Let $\hat H_M$ be an operator depending on a parameter $M$. We say that the $M$ dependent operator  $\hat{\mathcal{S}}_M$ is a \emph{shift operator} for $\hat{H}_M$ if there exists a
function $\bar M=\bar{M}(M)$  such that
	$$
	\hat{\mathcal{S}}_M \hat H_M=\hat H_{\bar{M}}\hat{\mathcal{S}}_M.
	$$ 
Let $\hat H$ be an operator and the function $\psi_\lambda$ be an eigenfunction of $\hat{H}$  with eigenvalue $\lambda$. We say that $\hat{\mathcal{L}}_\lambda$ is a \emph{ladder operator} for the eigenfunctions of $\hat H$ if there exists a function
 $\bar{\lambda}=\bar{\lambda}(\lambda)$ such that
$$
\hat H \psi_\lambda =\lambda \psi_\lambda \ \Rightarrow \ 
\hat{H}(\hat{\mathcal{L}}_\lambda \psi_\lambda) =  \bar{\lambda} \hat{\mathcal{L}}_\lambda\psi_{\lambda}.
$$ 
If $\bar{\lambda}(\lambda)=\lambda$ for all eigenvalues $\lambda$ we say that $\hat{\mathcal{L}}_\lambda$ is a \emph{symmetry} for
$\hat{H}.$	\end{defi}

We consider the Schr\"odinger operator associated with the natural Hamiltonian $L$ (\ref{L}), 
\begin{equation}\label{hL}
\hat L=-\frac {\hbar^2}2\Delta +V. 
\end{equation}
where $\Delta$ is the Laplace-Beltrami operator associated to the metric $\mathbf g$.
Let $\mathbf{\bar g}$ be  the
metric tensor of the extension of (\ref{L})  which is of the form
\begin{equation}
\bar g^{ij}=\left(\begin{matrix} 1 & 0 \cr 0 & \alpha(u)g^{ij} \end{matrix} \right),\quad \alpha(u)=\frac {m^2}{n^2}\left(c\gamma^2+{C}\right)=-k^2\gamma',
\end{equation}
since $\gamma$ satisfies (\ref{eqgam}). 
We choose $\mathbf{\bar g}$ as metric tensor  for the quantum extended Hamiltonian of (\ref{hL}).
Therefore, the Laplace-Beltrami operator $\bar \Delta$ associated with the metric $\mathbf{\bar g}$ is
\begin{equation}
\bar \Delta \psi=\partial ^2_u \psi+ N c\gamma \partial _u \psi-
k^2\gamma' \Delta \psi,
\end{equation}
where $N$ is the dimension of the base manifold of $L$.
We consider an extended Schr\"odinger operator of
$(\ref{hL})$ of the form
\begin{equation}\label{HHH}
\hat H=-\frac{\hbar^2}2 (\partial ^2_u \psi+ N c\gamma \partial _u \psi-
k^2\gamma' \Delta \psi) -k^2\gamma' V+ W(u), \end{equation}
where $k$ is a constant and the function $W$ will be suitably determined later.
If we consider eigenfunctions $\psi_\lambda$ of the operator $\hat{L}$
$$
\hat L\psi_\lambda=\lambda \psi_\lambda,
$$
then, for any $\phi=\phi(u)$,
$$
\hat H(\phi\psi_\lambda) =-\frac{\hbar^2}2\left(\partial_u^2 + Nc \gamma \partial_u\right)(\phi)\psi_\lambda+
(W-k^2\gamma'\lambda)\phi\psi_\lambda.
$$ 
Since by (\ref{eqgam}) $\gamma'=-c\gamma^2-C$, we may write
\begin{equation}\label{H1}
\hat H(\phi\psi_\lambda) =-\frac{\hbar^2}2\left(\partial_u^2 + Nc \gamma \partial_u\right)(\phi)\psi_\lambda +
(W+ck^2\lambda\gamma^2+ C k^2\lambda)
\phi\psi_\lambda.
\end{equation}
Hence, we have that $\hat{H}(\phi\psi_\lambda)=
E\phi\psi_\lambda$
(that is $\phi\psi_\lambda$ is an eigenfunction of (\ref{HHH}) associated with the eigenvalue $E$) if and only if $\phi(u)$ is 
an eigenfunction of the 
$M$-dependent operator 
\begin{equation}\label{hHM}
\hat{H}^M=
 -\frac{\hbar^2}2\left(\partial_u^2 + Nc \gamma(u) \partial_u\right) 
+M(c\gamma^2(u)+C)+W(u), 
\end{equation}
with $M=k^2\lambda$. 
We denote such a $\phi$ by $\phi^{k^2\lambda}_E$ in order to underline its dependence on the spectral parameters $E$ and $\lambda$. 

\begin{defi}  \label{d:ws}
We say that $\hat{X}$ is a \emph{warped symmetry operator} of 
(\ref{HHH}) if, for all partially separated eigenfunctions $f_E=\phi^{k^2{\lambda}}_E\psi_\lambda$, $\hat{H}\hat{X}(f_E)=E\hat{X}(f_E)$, or equivalently, that $[\hat{H},\hat{X}](f_E)=0$.
We allow $\hat{X}$ to depend on both $\lambda$ and 
$E$.
\end{defi}

In the following, we construct operators $\hat G$, $\hat A$, $\hat D$ which are the factors of a warped symmetry $\hat X$ of $\hat H$ for $k\in \mathbb{Q}$.
In particular, with Theorem \ref{teoq} below, we produce a quantum version of the factorized characteristic first integrals of extended Hamiltonians of Proposition \ref{teo}, valid only for the case $c\neq 0$, $C=0$.

Given any function $G(q^1,\ldots,q^N)$, we define
the first-order operators \begin{equation}\label{dhG}
\hat G^\pm_\epsilon=\aa \nabla^iG \nabla_i+a_1^\pm G+a_2^\pm
\end{equation}
where the constants $a_i$ verify the relations
\begin{equation}\label{be}
 a_1^\pm=\aa\left(\frac {c(1-N)}2\pm \epsilon \frac {\sqrt{2|c|}}\hbar\right), \qquad
a_2^\pm=\frac{-2c_1\aa}{-c\hbar^2\pm 2\hbar\epsilon\sqrt{2|c|}},\end{equation}
with $\epsilon, c_1\in \mathbb{R}$, $c\neq 0$.
Moreover, for any integer $n>0$, we construct the operators
$(\hat{G}^{\pm}_\epsilon)^n$ as follows
\begin{equation}\label{hGn}
(\hat{G}^{\pm}_\epsilon)^n= \hat{G}^{\pm }_{\epsilon\pm(n-1)s}\circ \hat{G}^{\pm }_{\epsilon \pm(n-2)s}\circ \cdots \circ \hat{G}^{\pm }_{\epsilon \pm s}\circ \hat{G}^{\pm }_{\epsilon} 
\end{equation}
where $s=\sqrt{\frac{|c|}{2}}\hbar,$ and $\hat{G}^\pm_\epsilon$ is defined in (\ref{dhG}).

We define the first-order operators
\begin{equation}\label{dhA}
\hat{A}^{\sigma,\tau}_\mu= \partial_u+\frac{1}{\hbar} \left[\sigma  {\omega} cu+\frac{\hbar c(N-1)+\tau\sqrt{2|c|}\mu}{2cu}
\right], \qquad (\sigma,\tau=\pm 1),
\end{equation}
\begin{equation}\label{Dhat}
\hat D_E^\pm= \frac {\hbar}{\sqrt{2}}u\partial_u
+\frac{\hbar (N+1)}{2\sqrt{2}}\pm\left( \frac{E}{2c\omega}-\frac{\omega}{\sqrt{2}} cu^2\right),
\end{equation}
where $E$, $\mu$, $\omega\in \mathbb{R}$, $c\neq 0$.
Moreover, for any integer $m>0$, we define the operators 
\begin{equation}\label{dAm}
(\hat{A}^{\sigma,\tau}_\mu)^m=\hat{A}^{\sigma,\tau}_{\mu +\tau(m-1)s}\circ \hat{A}^{\sigma,\tau}_{\mu + \tau(m-2)s}\circ \cdots \circ \hat{A}^{\sigma,\tau}_{\mu + \tau s}\circ \hat{A}^{\sigma,\tau}_{\mu}, 
\end{equation}
\begin{equation}\label{dDm}
(\hat{D}^{\pm}_E)^m=\hat{D}^{\pm }_{E \pm 2(m-1)\delta} \circ \hat{D}^{\pm }_{E \pm 2(m-2)\delta}\circ \cdots \circ \hat{D}^{\pm}_{E \pm 2\delta}\circ \hat{D}^{\pm}_{E}, 
\end{equation}
with $s=\sqrt{\frac{|c|}{2}}\hbar$, $\delta=\hbar c\omega,$  $\sigma,\tau=\pm 1$.
 
 In (\ref{hGn}) and (\ref{dDm}), the signs in upper
and lower indices must be the same (all + or all $-$).

We are able to state the theorem as follows 

\begin{teo}\label{teoq}
Let $L_0=\frac{1}{2}g^{ij}p_ip_j+V_0(q^i)$ be a classical natural Hamiltonian with $N$ degrees of freedom 
and $G$ a function of $(q^i)$ satisfying (\ref{HE1}), (\ref{VE1}) for  $c\neq 0$ and $c_0=0$
i.e., generating 
 a ladder function of the form (\ref{18}) for $L_0$.
For any $k\in \mathbb{Q}^+$, let us construct the polynomial differential operator
 \begin{equation}\label{hX}
 \hat{X}^k_{\epsilon,E}=(\hat{G}^{+}_\epsilon)^{2n} \circ
 (\hat{A}^{1,1}_{k\epsilon})^{2m}\circ (\hat{D}^{+}_E)^m
 \end{equation}
 where $m$,$n$ are positive integers  such that $k=m/n$, the operators
 $(\hat{G}^{+ }_\epsilon)^{2n}$, $(\hat{A}^{\sigma,\tau}_\mu)^{2m}$ (with $\mu=k\epsilon$, $\sigma=\tau=1$) and $(\hat{D}^{+}_E)^m$ are defined in (\ref{hGn})
 (\ref{dAm}) (\ref{dDm}) and 
 where $\hat{G}_\epsilon^\pm$ are constructed from the function $G$.
 Then, the
 quantum Hamiltonian 
\begin{equation}\label{hL0}
\hat{L}_0= -\frac{\hbar^2}2\Delta+V_0,
\end{equation}
associated with $L_0$, can be extended 
to the quantum Hamiltonian
\begin{equation}\label{Hokk}
\hat{H}=-\frac{\hbar^2}{2}\left(\partial^2_{uu}+ \frac{N}{u}\partial_u
\right)+\frac{k^2}{cu^2}\hat{L}_0 + \frac{\omega^2}{2}c^2u^2+ \frac{\dok (k^2+1)}{cu^2},
\end{equation}
with
\begin{equation}\label{dok}
\dok=-\frac{\hbar^2c(N-1)^2}{8},
\end{equation}
which admits the operator $\hat{X}^k_{\epsilon, E}$ as warped symmetry: i.e., 
\begin{equation*}\label{hHhX}
\hat{H}(\hat{X}^k_{\epsilon, E} f_E)= E\hat{X}^k_{\epsilon, E} (f_E),
\end{equation*}
for all (partially) multiplicatively separated eigenfunctions $f_E=\phi_E^{M}(u)\psi_\lambda(q^i)$ of $\hat{H}$ such that $\psi_\lambda$ is an eigenfunction of $\hat{L}_0$ with eigenvalue $\lambda$,  and
$$M=k^2(\lambda+\dok),\qquad \epsilon
=\sqrt{| \lambda +\dok|}, \qquad
c(\lambda+\dok)\ge 0.$$
\end{teo}

\begin{rmk}

For $N=1$, the quantum extended Hamiltonian (\ref{Hokk}) is
 the quantum operator associated with the
extension of $L_0$ with $\gamma=1/(cu)$ (that is for $c\neq0$, $C=0$).
However, for $N>1$ (that is when $L_0$ has more than one degree of freedom) this is no more true, due to 
 the presence of the last 
addendum in (\ref{Hokk}). 
The addendum could be absorbed in $L_0$, by replacing $L_0$ with
${L}_1=L_0+\dok\frac{k^2+1}{k^2}$.
Nevertheless, in the proof of Theorem \ref{teoq} it is
essential to consider the eigenvalues of $\hat{L}_0$, not those of the shifted operator $\hat{{L}}_1$. 
From a different view point, the additional term proportional to $\dok$  added to
the quantization of the classical extension of $L_0$
suggests a link with the introduction of a kind of quantum correction, used as instance in \cite{CDRmlbq, B2} to allow quantization for classical quadratic in the momenta first integrals.
\end{rmk}

In \cite{CRsl} it is shown that this form of quantization recovers the quantization of the TTW system shown in \cite{KN}, generalizing the approach to N dimensional manifolds, in the case of any extension for $N=1$, and for the extensions of any $N$ dimensional Hamiltonian $L$ on non-flat constant curvature manifolds.

For the quantum case, it is easy to see that the degeneration of the energy levels, associated with the superintegrability of quantum systems, is affected by certain values of $k$. In \cite{Prague}, the degeneration of the energy levels of the Kepler-Coulomb system is shown to be less than maximal for certain values of $k$, when the system is obtained from (\ref{PW1}) by putting $c_1=c_2=0$. The same happens for the Klein-Gordon equation on Schwarzschild background.

\section{Conclusions}

In this review, we summarized the results of several years of our research about superintegrable systems appeared in a number of articles. This allows to put them in to a unified perspective, enhancing the underlying geometric structures of warped manifolds and Riemannian coverings. We may conclude that the whole procedure of extension of Hamiltonian systems, allowing the construction of new superintegrable systems from know ones with non trivial first integrals of arbitrarily high degree, is made possible by the warped structure of the metric of the given configuration space, and it ends on Riemannian coverings  of any rational order of that space.  The knowledge of these geometric structures puts in evidence issues of global definition of some of the first integrals. In several cases, we developed a quantization procedure to obtain symmetry operators from the classical first integrals, procedure that, for the moment, is not applicable to all extended Hamiltonians. Our aim  with this review is to ease the approach to our work on superintegrability,   allowing its application  by the scientific community.  


\section{Conflict of interest}
The authors declare no conflict of interest.


\begin{thebibliography}{999}

\bibitem{MPW} W. Miller Jr., S. Post, P. Winternitz, Classical and quantum superintegrability with applications, {\it J. Phys. A: Math. Theor.} {\bf 46} (2013), 423001, 97 pages

\bibitem{BCRCal}  S. Benenti, C. Chanu, G. Rastelli, The super-separability of the three-body inverse-square Calogero system, {\it J. Math. Phys.} {\bf 41} (2000), 4654–4678 

	
\bibitem{s1} J. Kress, K. Schöbel, A. Vollmer, Algebraic Conditions for Conformal Superintegrability in Arbitrary Dimension. {\it Commun. Math. Phys.} {\bf 405}, 92 (2024). https://doi.org/10.1007/s00220-023-04872-w

\bibitem{s2}   R. Campoamor-Stursberg, On some algebraic formulations within universal enveloping algebras related to superintegrability, {\it Analytic and Algebraic Methods in Physics 
} {\bf 62} No. 1 (2022)

\bibitem{s3} A. Najafizade,
 H. Panahi, Classical and quantum dynamics of a constrained particle on three-dimensional spaces of constant curvature: An algebraic approach on the superintegrable problem, {\it Modern Physics Letters A} {\bf 37}, No. 22, (2022) 2250140  https://doi.org/10.1142/S0217732322501401],

\bibitem{s4} D. R. Nozaleda
, P. Tempesta, G. Tondo, Classical multiseparable Hamiltonian systems, superintegrability and Haantjes geometry, {\it Communications in Nonlinear Science and Numerical Simulation} {\bf 104} (2022) 106021

\bibitem{s5} M. Błaszak, K. Marciniak Classical and Quantum Superintegrability of Stäckel Systems, {\it SIGMA} {\bf 13 }(2017), 008, 23 pages      arXiv:1608.04546      https://doi.org/10.3842/SIGMA.2017.008

\bibitem{s6} E. G. Kalnins, J. M. Kress, W. Miller Jr, Separation of Variables and Superintegrability
The symmetry of solvable systems, {\it IOP Publishing Ltd} (2018)
Online ISBN: 978-0-7503-1314-8 • Print ISBN: 978-0-7503-1315-5 

\bibitem{s7} A.V. Tsiganov, Transformation of the Stäckel matrices preserving superintegrability, {\it J. Math. Phys.} {\bf 60}, 042701 (2019)
https://doi.org/10.1063/1.5057885

\bibitem{s8} C. Gonera, M. Kaszubska, Superintegrable systems on spaces of constant curvature, {\it Annals of Physics} {\bf 346} (2014), Pages 91-102

\bibitem{s9} A. P. Fordy, Q. Huang, Superintegrable systems on 3 dimensional conformally flat spaces, {\it Journal of Geometry and Physics} {\bf 153}, (2020) 103687

\bibitem{s10} J. F. Cariñena, M. F. Rañada, M. Santander Superintegrability of three-dimensional Hamiltonian systems with conformally Euclidean metrics. Oscillator-related and Kepler-related systems, {\it J. Phys. A: Math. Theor.} {\bf 54} (2021) 105201

\bibitem{s11} J. F. Cariñena, F. J. Herranz, M. F. Rañada, Superintegrable systems on 3-dimensional curved spaces: Eisenhart formalism and separability,
{\it J. Math. Phys.} {\bf 58}, 022701 (2017)
https://doi.org/10.1063/1.4975339

\bibitem{s12} J. Kress, K. Schöbel, A. Vollmer, Superintegrable systems on conformal surfaces, 	{\it arXiv:2403.09191} [math.DG] (2024) https://doi.org/10.48550/arXiv.2403.0919]

\bibitem{s13} G. Valent, Superintegrability on the hyperbolic plane with integrals of any degree $\geq 2$, {\it Journal of Geometry and Physics}
{\bf 183}, ( 2023), 104686


\bibitem{s14} M. F. Rañada, Superintegrable deformations of superintegrable systems: Quadratic superintegrability and higher-order superintegrability, {\it J. Math. Phys.} {\bf 56}, (2015) 042703


\bibitem{s15} I. Marquette, P. Winternitz,  Higher Order Quantum Superintegrability: A New “Painlevé Conjecture”. {\it In: Kuru, Ş., Negro, J., Nieto, L. (eds) Integrability, Supersymmetry and Coherent States. CRM Series in Mathematical Physics. Springer, Cham.} (2019) https://doi.org/10.1007/978-3-030-20087-9 4

\bibitem{Bo}   A. V. Borisov,  A. A. Kilin, I. S. Mamaev, Superintegrable system on a sphere with the integral of higher degree, {\it Regul. Chaotic Dyn.} {\bf 14}  no. 6 (2009), 615-620 


\bibitem{s16} O. Kubů,
, A. Marchesiello, L. Šnobl, Superintegrability of separable systems with magnetic field: the cylindrical case,  {\it J. Phys. A: Math. Theor.} {\bf 54} (2021) 425204

\bibitem{s17} A. Marchesiello, D. Reyes, L. Šnobl, Superintegrable families of magnetic monopoles with non-radial potential in curved background, {\it Journal of Geometry and Physics}
{\bf 203}, (2024) 105261








   







\bibitem{CDR0}   C. Chanu,  L. Degiovanni, G. Rastelli, Superintegrable three-body systems on the line,  \textit{J. Math. Phys.} {\bf 49}, (2008) 112901 

\bibitem{TTW}  F. Tremblay,   A.V. Turbiner, P. Winternitz, An infinite family of solvable and integrable quantum systems on a plane, \textit{J. Phys. A} {\bf 42}  no. 24, (2009) 242001, 10 pp. 

\bibitem{TTWcdr}  C. Chanu,  L. Degiovanni, G. Rastelli, The Tremblay-Turbiner-Winternitz system as extended Hamiltonian, {\it J. Math. Phys.} {\bf 55} (2014) 122701 


\bibitem{KM1} E. G. Kalnins, J.  Kress, W.  Miller Jr, Superintegrability and higher order integrals
for quantum systems, {\it J. Phys. A: Math. Theor.} {\bf 43} (2010) 265205

\bibitem{PWccm} S. Post, P. Winternitz,  An infinite family of superintegrable deformations of the Coulomb potential,
{\it J. Phys. A: Math. Theor.} {\bf 43}, (2010) 222001, 11 pages

\bibitem{KMccm}  E. G. Kalnins , W.  Miller Jr, S. Post, Coupling constant metamorphosis and Nth-order symmetries in classical
and quantum mechanics, {\it J. Phys. A: Math. Theor.} {\bf 43}, (2010) 035202, 20 pages

\bibitem{CDRPol} C. Chanu,  L. Degiovanni, G. Rastelli, Polynomial constants of motion for Calogero-type systems in three dimensions, \textit{J. Math. Phys.} \textbf{52}, (2011) 032903 
	
\bibitem{CDRraz} C. Chanu,  L. Degiovanni, G. Rastelli, Extensions of Hamiltonian systems dependent on a rational parameter, {\it J. Math. Phys.} {\bf 55} (2014) 122703


\bibitem{CDRpw}  C. Chanu,  L. Degiovanni,  G. Rastelli, Extended Hamiltonians, coupling-constant metamorphosis and the Post-Winternitz syste,  {\it SIGMA} {\bf 11}  094, (2015) 9 pp. 

\bibitem{Rbj} G. Rastelli, Born-Jordan and Weyl Quantizations of the 2D Anisotropic Harmonic Oscillator, {\it SIGMA} {\bf 12} (2016), 081, 7 pages

\bibitem{CDRmlbq}  C. Chanu,  L. Degiovanni, G. Rastelli, Modified Laplace-Beltrami quantization of
	natural Hamiltonian systems with quadratic
	constants of motion,  {\it J. Math. Phys.} {\bf 58} (2017) 033509 

\bibitem{CRsl} C. M. Chanu, G.  Rastelli, Extended Hamiltonians and shift, ladder
functions and operators, {\it Annals of Physics} {\bf 386}, (2017) 254-274

\bibitem{CDRfi}  C. Chanu,  L. Degiovanni, G. Rastelli,  First integrals of extended Hamiltonians in $(n+1)$-dimensions generated by powers of an operator,  {\it SIGMA} {\bf 7}  (2011) 038, 12 pp. 

\bibitem{CDRgen} C. Chanu,  L. Degiovanni, G. Rastelli, Generalizations of a method for constructing first integrals  of a class of natural Hamiltonians and some remarks about quantization,
	\textit{J. Phys.: Conf. Ser.} \textbf{343} (2012) 012101  

\bibitem{coverings}  C. M.  Chanu, G. Rastelli, Separation of variables and superintegrability on Riemannian coverings, 	{\it SIGMA} {\bf 19} (2023), 062, 18 pages 


\bibitem{7ter}  M. F. Ra\~{n}ada, M. Santander, Superintegrable systems on the two-dimensional sphere S2 and the hyperbolic plane H2, {\it J. Math. Phys.} {\bf 40}, (1999) 5026-5057

\bibitem{Ta}  Y. Tashiro,  Complete Riemannian manifolds and some vector fields, {\it Trans. Amer. Math. Soc.} {\bf 117} (1965) 251-275

\bibitem{CDRGhent}  C. Chanu, L. Degiovanni, G. Rastelli, Warped product of Hamiltonians and extensions of Hamiltonian systems, {\it Journal of Physics: Conference Series} {\bf 597}, (2015) 012024


\bibitem{CRbs}  C. M. Chanu, G. Rastelli, Block-Separation of Variables: a Form of Partial Separation for Natural Hamiltonians, {\it SIGMA} {\bf 15} (2019), 013, 22 pages 

\bibitem{CDRsuext} C. Chanu,  L. Degiovanni,  G. Rastelli,  Superintegrable extensions of superintegrable systems,  {\it SIGMA} {\bf 8}  070, (2012) 12 pp. 

\bibitem{NoNN} C. M. Chanu, G. Rastelli,  Extensions of nonnatural Hamiltonians, {\it Theor. Math. Phys.}  {\bf 204}, 1101–1109 (2020). https://doi.org/10.1134/S0040577920090019


\bibitem{CRcs}  C. M. Chanu, G. Rastelli, On the Extended-Hamiltonian Structure of Certain Superintegrable Systems on Constant-Curvature Riemannian and Pseudo-Riemannian Surfaces, 	{\it SIGMA} {\bf 16} (2020), 052, 16 pages

\bibitem{CS}  R. Campoamor-Stursberg,  Superposition of super-integrable pseudo-Euclidean potentials in N=2 with a fundamental constant of motion of arbitrary order in the momenta, {\it J. Math. Phys.} {\bf 55}, (2014) 042904

\bibitem{Hi}  J. Hietarinta, B. Grammaticos, B. Dorizzi, A. Ramani, Coupling-constant metamorphosis and duality between integrable Hamiltonian systems, {\it Phys. Rev. Lett.} {\bf 53}, (1984) 1707-1710 

\bibitem{CKNd}  E. Celeghini,  S. Kuru,  J. Negro, M.A. del Olmo, A unified approach to quantum and classical TTW systems based on factorization, {\it Ann. Phys.} 332, (2013) 27-37 

\bibitem{KN}  S. Kuru, J. Negro,  Factorizations of one-dimensional classical systems, {\it Ann. Phys.} {\bf 323}, (2008) 413-431

\bibitem{B2}  A. Ballesteros,  A. Enciso,  F. J. Herranz,  O. Ragnisco,   D. Riglioni, Quantum mechanics on spaces of nonconstant
	curvature: The oscillator problem and superintegrability, \textit{Ann. Phys.} \textbf{326} (8), (2011) 2053-2073 
	
		

    \bibitem{Prague} C. M. Chanu, G. Rastelli, On the degeneracy of the energy levels of Schr\"odinger and Klein-Gordon equations on Riemannian coverings, {\it arXiv:2411.10824}, (2024)

    
	
	
	
	
	
	
	
	
	
	
		
	
	


























%
%
%
%
%
%
%
%
%
%
%























\end{thebibliography}
\end{document}